\RequirePackage{fix-cm}
\documentclass[smallextended]{svjour3}       
\smartqed  
\usepackage{graphicx}
\usepackage{caption}

\usepackage{url}

\usepackage[T1]{fontenc}
\usepackage{tabularx}
\usepackage{gensymb}
\usepackage{siunitx}
\usepackage{units}
\usepackage{tabu}
\usepackage[dvipsnames]{xcolor}
\usepackage{amsmath}
\usepackage{alphabeta}
\begin{document}

\title{Light extraction from CVD-grown <400> single crystal diamond nanopillars. Selective charge state manipulations with 0V \textit{SF\textsubscript{6}} plasma.}

\titlerunning{Light extraction from CVD-grown <400> single crystal diamond nanopillars.} 

\author{Mariusz Radtke\textsuperscript{a} \and Abdallah Slablab \textsuperscript{b} \and Sandra Van Vlierberghe\textsuperscript{a}      \and Chao-Nan Lin \textsuperscript{c} \and Ying-Jie Lu \textsuperscript{c} \and Shan Chong-Xin\textsuperscript{c} }

\institute{\textsuperscript{a}Ghent University, Department of Organic and Macromolecular Chemistry, PBM, CMaC  \at
              9000, Ghent, Belgium \\
              \and \textsuperscript{b}Saarland University, Department of Physics \at 66123 Saarbrücken, Germany \and \textsuperscript{c}Zhengzhou University, School of Physics and Microelectronics, Henan Key Laboratory of Diamond Optoelectronic Materials and Devices \at Zhengzhou 450001, China\\
              \email{M. Radtke, mariusz.radtke@ugent.be}           
           \and
           C.X. Shan \at
               cxshan@zzu.edu.cn
}

\date{Received: date / Accepted: date}

\maketitle

\begin{abstract}
We investigate the possibilities to realize light extraction from single crystal diamond (SCD) nanopillars. This was achieved by dedicated 519 nm laser-induced spin-state initiation of negatively charged nitrogen vacancies \textit{(NV\textsuperscript{-})}. For the first time, we present possibility to perform effective spin-readout of \textit{NV\textsuperscript{-}}s that were naturally generated by the growth process during chemical vapor deposition (CVD) synthesis within SCD without any post-growth implantation strategies. Applied diamond was neither implanted with \textsuperscript{14}N\textsuperscript{+}, nor was the CVD synthesized SCD annealed, making the presence of nitrogen vacancy a remarkable phenomenon. To investigate the possibility to realize light extraction by the utilization of \textit{NV\textsuperscript{-}} bright photoluminescence at room temperature and ambient conditions with the waveguiding effect, we have performed a top-down nanofabrication of SCD by electron beam lithography (EBL) and dry inductively-coupled plasma/ reactive ion etching (ICP-RIE) to generate light focusing nanopillars. In addition, we have fluorinated the diamond's surface by dedicated 0V  ICP plasma. Light extraction and spin manipulations were performed with photoluminescence (PL) spectroscopy and optically detected magnetic resonance (ODMR) at room temperature. We have observed a remarkable effect based on the selective 0V SF\textsubscript{6} plasma etching and surprisingly, in contrast to literature findings, deactivation of \textit{NV\textsuperscript{-}} centers.We discuss the possible deactivation mechanism in detail regarding 2-dimensional hole gas (2HG) and Fermi band bending.
\end{abstract}

\section{Introduction}
\label{intro}
Diamond is the hardest material in Mohs scale that attracts attention spanning from antipodial fields of industrial abrasives to even quantum information \cite{ettore2017}. In theory, diamond is composed only of isotopic mixture of sp\textsuperscript{3} carbon atoms (\textsuperscript{12}C and \textsuperscript{13}C), which are ideally aligned in fused, chair conformed megastructures, which build up the diamond supercell \cite{PhysRevLett.120.136401}. However, this idealized imagination of diamond is not complete, as in reality within the supercell not every carbon atom contains saturated bonds, mostly due to the impurities and crystallographic imperfections incorporated into lattice during the growth \cite{B908532G}. The most common impurity is nitrogen (\textsuperscript{14}N), which is vastly present in synthetic diamond, giving it a remarkable yellow appearance \cite{chakraborty2019}. The nitrogen content among other impurities is basis for its classification ranging from Type 1b to optical and electronic grade. The most common synthetic methods of diamond growth are high pressure high temperature (HPHT) for Types 1a and 1b and chemical vapor deposition (CVD) for Type 2b and electronic grade.  As chemical vapor deposition occurs at elevated temperatures reaching up to 1000 $\deg$C and high vacuum reaching 10E-06 mTorr, the deposited layers are annealed in those conditions \cite{osterkamp2019}. As nitrogen being the main impurity within the diamond supercell in its sp\textsuperscript{3} is able to only utilize 3 covalent bonds and due to the fact that carbon (sp\textsuperscript{3}) can have 4 bonds, the missing gap after incorporation of nitrogen into supercell results in so-called nitrogen vacancy, \textit{NV\textsuperscript{-}} \cite{PhysRevB.86.125204}. This vacancy has remarkable properties. It behaves like an artificial atom, has a well-defined electronic strucuture that can be simulated by sophisticated methods of density functional theory (DFT) and most importantly from the quantum sensing perspective- it has an electron spin (S=1) \cite{PhysRevLett.101.117601}. The presence of electron spin causes its instrinsic angular momentum to couple to a laser at resonance frequency, causing an excitation and following relaxation resulting in bright photoluminescence already at room temperature. A nitrogen vacancy comes in many different forms, whereby the negatively charged nitrogen vacancy draws most of the attention in quantum-related research field. The reason for that arises from the fact that \textit{NV\textsuperscript{-}} has 6 electrons in its structure contributing to the electron spin S=1 \cite{gali2019}. This electron spin can be optically excited and the resulting fluorescence at room temperature is bright enough to be easily detected by a charge coupled device (CCD) of a standard confocal microscopy. Due to the presence of electron spin, the negatively charged nitrogen vacancy can also couple to the external magnetic field, which influences its fluorescence upon appropriate laser excitation. This allows the use of diamond in magnetometry, metrology and also quantum computing \cite{radtkenanoscale2019}. All of that is possible as diamond is a material with neglegible optical absorption in the region above 220 nm up to even 1000 $\mu$m due to its band gap of approximate 5.5 eV (with the exception of a weak absorption line at 5 $\mu$m due to the lattice absorption resulting in transformation of photons into phonons) \cite{mildren2013}. The enhancement of photoluminescence count-rate arising from \textit{NV\textsuperscript{-}} relaxation can be achieved with waveguiding effect. The easiest approach to do so is nanofabrication \cite{radtkereliable2019} to focus the light leaving diamond in principle similar to an optical lense. Diamond is nevertheless an insulator, which makes it a challenging material to perform standard lithogrpahy by means of electron beam (EBL) and use of reactive ions in plasma to etch it (ICP/RIE), mostly due to the capactive charging effects. These nanofabrication issues can be resolved by sophisticated use of decharging layers and dedicated plasma technology \cite{radtkereliable2019}. For the efficient utilization of waveguiding effect, the diamond surface quality is of paramount importance. Imperfections of square mean roughness may cause the photoluminescence signal to be diffracted from the surface imperfections of roughness exceeding 1 nm, which leads to serious signal loss \cite{favaro2015}. Other method to increase the photoluminescence count-rate and stability of \textit{NV\textsuperscript{-}} in diamond is the use of dedicated surface termination. The most common cause of drop in photoluminescence signal from \textit{NV\textsuperscript{-}} is loss of one electron from its intrinsic electronic structure due to the photoelectric effect induced by the incident laser. The ejection of an electron from the valence band to conduction band causes the transition of \textit{NV\textsuperscript{-}} into the optically dormant \textit{NV\textsuperscript{0}} state with 5 electrons that cannot be optically read-out anymore. It has to be mentioned though that this undesired effect is also nowadays utilized for resolution-enhanced \textit{NV\textsuperscript{-}} detection \cite{bourgeoisphotoelectric2015}. As it was already described by others, this loss can be controlled by shifting the Fermi level between the valence and conduction band in diamond being separated by 5.5 eV energy gap \cite{grotzcharge2012-1}. The electron donating counterparts (hydrogen or methyl) cause the Fermi level to rise and therefore facilitate electron ejection upon laser excitation, see Fig.\ref{fig:1} . The use of electron-withdrawing groups like carboxyl, carbonyl or halogen groups has an opposite effect. The Fermi level drops, while the effective transition dipole moment of diamond changes its orientation and electron ejection is no more energetically unfavorable \cite{fengnanli2017}. This has an advantage for the nitrogen vacancy to retain its stability upon constant laser excitation. The most promising candidate for this process is fluorine termination. The currently investigated methods are wet chemical and plasma procedures \cite{hu2013}. Chemical fluorination does not cause surface roughening, but is on the other hand lenghty in time due to the chemical inertness and difficulties to ensure activation of C-H bond \cite{cui2013}. Plasma methodology is on the contrary very quick and convenient, yet can nevertheless cause surface roughening that leads to photoluminescence signal loss. The main parameter of plasma responsible for the physical and chemical etching character is the bias, which translates to acceleration of reactive ions towards the to-be etched surface \cite{Wakui2017}. The higher the value of bias, the more severe are the effect on the surface. Roughening occurs then mainly due to the sputtering of the resists or the surface itself as direct cause of physical etching. By lowering the bias even to 0V, the reactive ions will be swiftly surrounding the surface and react only in a chemical way \cite{mi10110718}. In the present study, we have developed a new type of 0V SF\textsubscript{6} plasma and have investigated its effect on CVD grown diamond with nitrogen vacancies generated by in-situ annealing during the growth process. Suprisingly, we have observed a selective deactivation of the naturally generated \textit{NV\textsuperscript{-}} caused by fluorination. We discuss this finding in terms of surface quality and introduce phenomenon occuring in the thermodynamic stability of the \textit{NV\textsuperscript{-}} generated by non-conventional natural growth.

\begin{center}

  \includegraphics[width=1.2\linewidth]{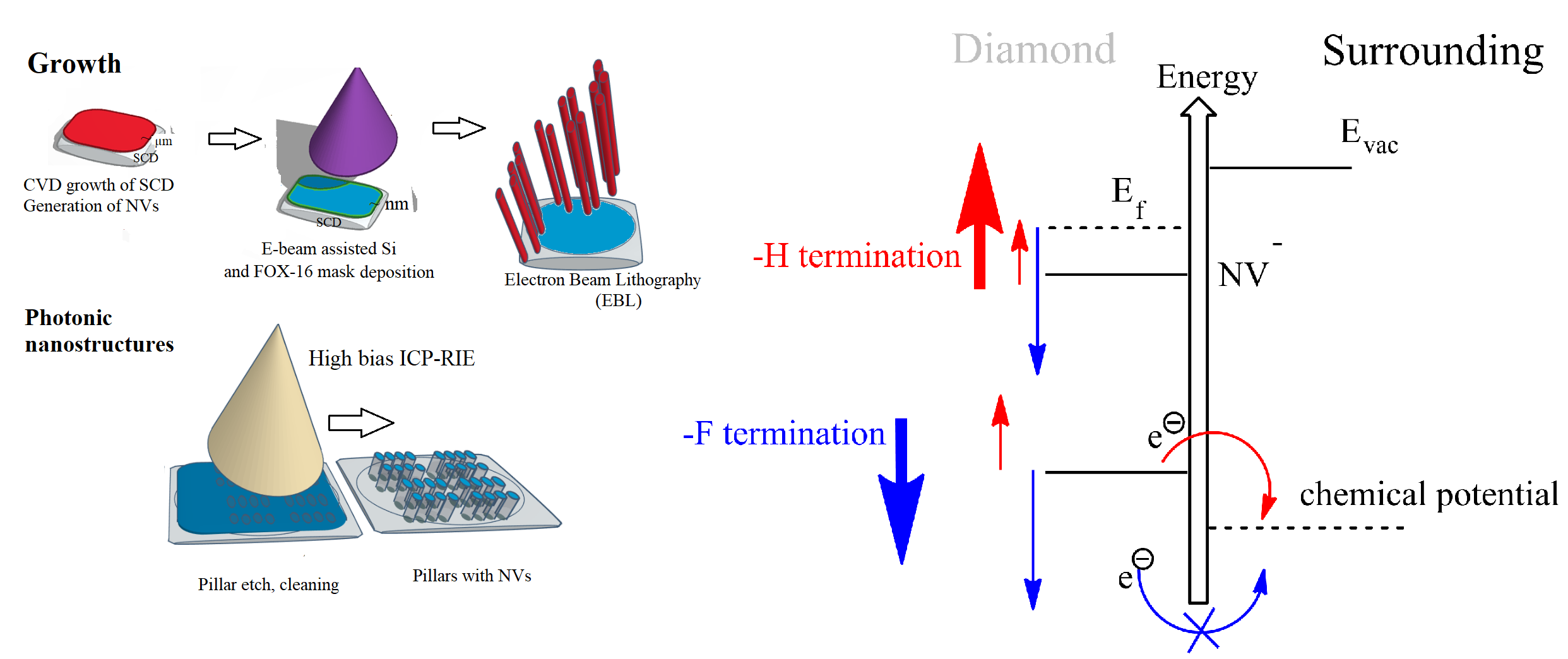}
\captionof{figure}{Left: generation of photonic nanostructures in CVD-grown single crystal diamond containing naturally grown-in, negatively charged nitrogen vacancies. Right: Effects of surface termination on the electron loss from a negatively charged nitrogen vacancy on example hydrogen and fluorine termination. In this study plasma fluorination with the use of heavy ions SF\textsubscript{6} was applied.}\label{fig:1}%
\end{center}

\section{Materials and Methods}
\label{sec:1}

\subsection{CVD diamond growth}
\label{sec:2}
Single crystal diamond was grown by chemical vapor deposition according to the procedure published earlier \cite{chen2018}. The presence of single crystal diamond was confirmed by the X-ray diffraction (XRD) measurement, which resulted with a distinctive peak at  119.8 $\deg$, which corresponds to <400> orientation \cite{C8TC04258F}.
\subsection{Nanofabrication}
\label{sec:3}
We refer the reader to earlier studies on the nanofabrication published earlier\cite{radtkereliable2019}. In brief: we employed a cold-cathode scanning electron microscope (SEM) (Hitachi S4500), equipped with RAITH Elphy software to perform electron beam-lithography (EBL). Immediately after the growth and acid cleaning of diamond, 25 nm thick polycrystalline silicon was evaporated on the surface as adhesive layer and a negative tone resist, FOX16 was spin-coated. The mask structures were written by electron beam lithography (\ref{fig:1}red). The written structures were further etched-in with highly anisotropic pure oxygen RIE/ICP plasma preceeded by a short SF\textsubscript{6} pulse. After removing mask and adhesive layer by wet-chemical etching methods (HF-based buffered oxide etchant and followed by immediate 3M KOH bath) as well as acid cleaning (1:1:1 HNO\textsubscript{3}, H\textsubscript{2}SO\textsubscript{4}, HClO\textsubscript{4}) the generated nanopillars were investigated by means of photoluminescence spectroscopy and optically detected magnetic resonance. We use FOX 16 (Dow, Corning) as a negative-tone resist, which after the exposure was developed in 25 $\%$ tetramethylammonium hydroxide (TMAH). The structures were dry etched without further purfication in an inductively coupled plasma/reactive ion etching (ICP/RIE) plasma reactor (Oxford PlasmaLab 100). 
\subsection{Photoluminescence spectroscopy on naturally grown-in nitrogen vacancies}
\label{sec:5}
A custom-built confocal microscope of 0.8 numerical aperture (NA) was used to excite and characterize negatively charged nitrogen vacancies in single crystal diamond. The confocal filtering was faciliated by a single mode fiber and the laser of choice was a continuous diode-pumped solid-state laser (DPSS) with a wavelength of 519 nm. The photoluminescence signal was detected through the application of a 650 nm longpass-filter and Eselitas SPCM-AQRH-14 photon counters. The spin manipulations were performed with a microwave source (Standford Research Systems, SG 384) and Mini Circuits amplifier, ZHL -42W+. The microwaves were delievered by 20]$\mu$m thick gold wire.
\section{Photonic properties of as-grown diamond and after successful nanofabrication}
\label{sec:6}
\begin{figure}

\centering
  \includegraphics[width=1.2\linewidth]{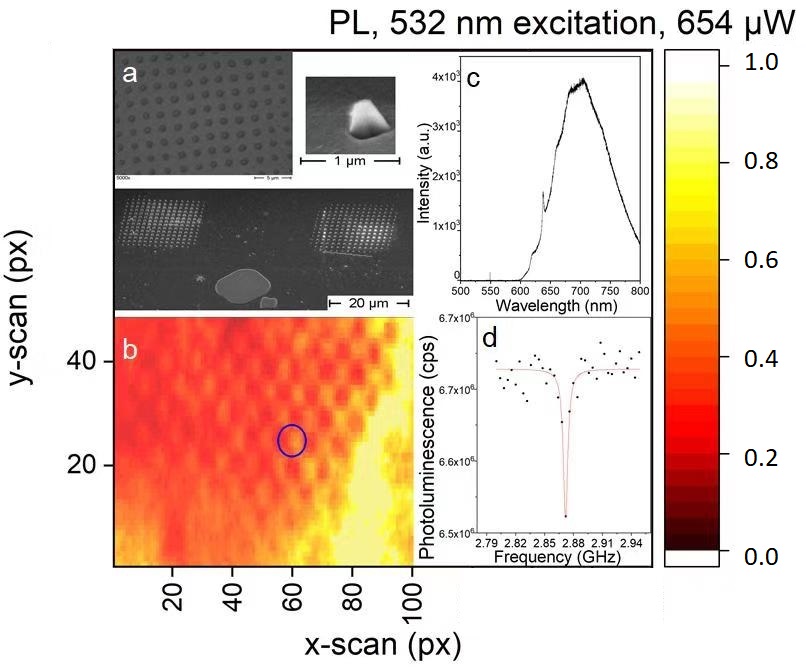}

\caption{Generation of diamond nanopillars on 2.3 cm thick CVD-grown diamond plate (a) and their photoluminescence mapping (b). Inset (c) represents photoluminescence spectrum with clear zero phonon line (ZPL) emerging from the presence of negatively charged nitrogen vacancy and Raman diamond line at 551 nm. Inset (c) shows optically detected magnetic resonance investigation with characteristic 2.87 GHz transition emerging from the degeneration of the m\textsubscript{s}=$\pm$1 state within nitrogen vacancy electronic structure with 2.8 $\%$ detection contrast.}
\label{fig:2}       
\end{figure}

Naturally grown-in nitrogen vacancies have shown a remarkable stability during irradiation. It has been obeserved by others that negatively charged nitrogen vacancies usually change their charge state from -1 to neutral and sometimes even +1 upon prolonged laser exposure \cite{bourgeoisphotoelectric2015}. This is caused by ejection of electrons from the valence to the conduction band. Negatively charged nitrogen vacancies are composed of 6 electrons that make up the unique properties of NVs. This electronic structure is responsible for the presence of the electronic spin angular momentum causing electronic dipole moment transition and resulting in optically detectableL photoluminescence. The detailed presentation of the electronic structure of nitrogen vacancies is shown in the figure \ref{fig:3}. In case of naturally grown nitrogen vacancies, we have observed the prolonged stability of the charge state, allowing for the multiple accumulation of photoluminescence maps and ODMR signals with and without applied magnetic field Fig.\ref{fig:2}. We note the fact that single application of the magnetic field induced a degeneration of m\textsubscript{s}=$\pm$1 state and a drop of the photoluminescence signal at the 2.87 GHz transition frequency, which usually requires multiple scans in order to obtain an improved signal to noise ratio. As the ground state of a nitrogen vacancy consists of a pair of ground and excited triplet and singlet states, the triplet ground state is made up  of m\textsubscript{s}=$\pm$1 and lower in energy  m\textsubscript{s}=0 Fig.\ref{fig:3}. As the $\pm$ ground state lies higher in energy, it is easier to excite it with e.g. a green 519 nm laser, which brings the nitrogen vacancy to its excited triplet state. There are three posssibilities for a relaxation here. First two possibilities occurs simultanously into 637 nm i.e. so-called purely electronic transition, zero phonon line (ZPL) to the m\textsubscript{s}=$\pm$1 and less energetically favored  m\textsubscript{s}=0 state \cite{C9TC01954E}. At room temperature these two transitions are seen as one ZPL, while at cryogenic temperatures, it is possible to observe two distinct lines \cite{C9TC01954E}. Indeed these transitions occur within one triplet system, they are symmetrically allowed. m\textsubscript{s}=$\pm$1 is therefore described as "bright state", main observed PL signal arises from this transition. There is also a possiblity of an other transition caused by intersystem crossing from the excited triplet state to the singlet state. As this transition is "forbidden" by symmetry rules, it occurs at significantly slower rates compared to the triplet-triplet transitions and the return to the m\textsubscript{s}=0 state is called a "dark" transition while m\textsubscript{s} is a "dark state" \cite{PhysRevB.98.075201}. In our case, we were able to distinguish between the "dark" and "bright" state of naturally grown-in, negatively charged nitrogen vacancies by means of photoluminescence spectroscopy. We observed a bright photoluminescence of zero phonon line in the nanofabricated diamond, even though it theoretically contributes only in 3$\%$ to the overall signal in Fig. \ref{fig:2}, and Fig.\ref{fig:3}. We were not only able to optically initialize the m\textsubscript{s}=$\pm$1 state, but also to manipulate it by an external magnetic field  in Fig. \ref{fig:2}. By sweeping the magnetic field around the photonic nanostructures, we observed a typical resonance at 2.87 GHz for negatively charged nitrogen vacancies. This resonance is typical for the  m\textsubscript{s}=$\pm$1 $\rightarrow$  m\textsubscript{s}=0 cycling transition. We note that by application of the external magnetic field, we observed a noisy splitting of the sharp singlet to a characteristic set of 4 doublets, indicating the m\textsubscript{s}=0 $\rightarrow$  m\textsubscript{s}=1 and m\textsubscript{s}=0 $\rightarrow$  m\textsubscript{s}=-1 transitions in all four possible crystallographic directions of \textit{NV\textsuperscript{-}s} \cite{PhysRevB.97.195448}.

\section{Plasma-aided manipulations of \textit{NV\textsuperscript{-}} charge state. The role of surface termination on \textit{NV\textsuperscript{-}}/N\textsuperscript{+} ratio.}

\begin{figure}

\centering
  \includegraphics[width=1.2\linewidth]{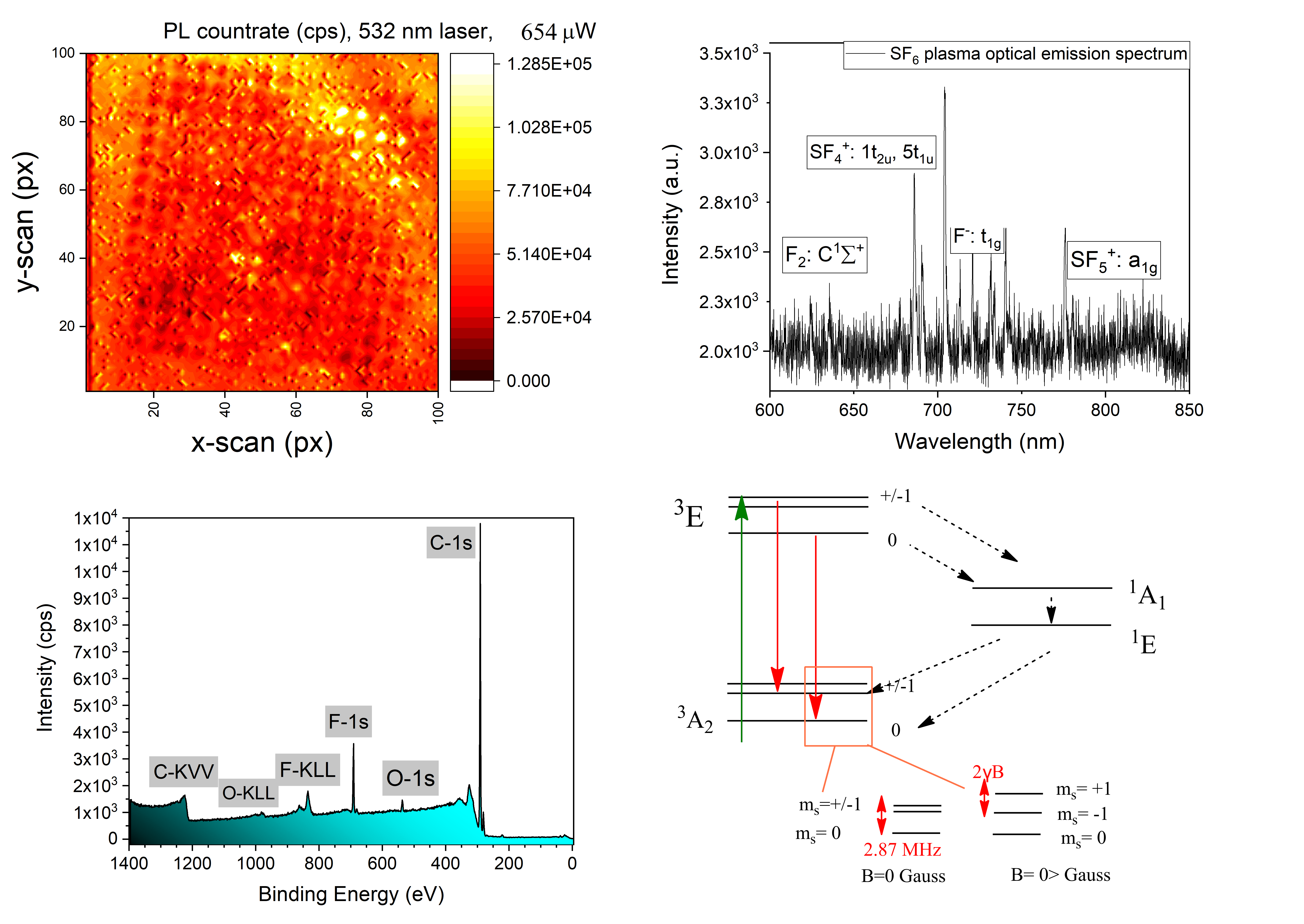}

\caption{(A) Photoluminescence scan of nanopillars etched into diamond after 0V bias plasma fluorination. The pillars were partially covered with single crystal quartz plate in order to shield them from the influence of plasma. Such an approach allows quantification of the fluorine termination on negatively charged nitroogen vacancies. An effect of photoluminescence quenching is clearly visible. (B) Optical emission spectrum of 0V bias SF\textsubscript{6} used to fluorinate the diamond with respective transitions. (C)X-ray photoelectron survey of the diamond fluorinated by SF\textsubscript{6} plasma with characteristic F-1s and F-KLL lines. (D) A detailed electronic structure with respective transitions within negatively charged nitrogen vacancies upon laser irradiation. }
\label{fig:3}   
\end{figure}

Inspired by the recent research performed on diamond surface modifications, we have attempted to fluorinate the surface of diamond to increase the stability of \textit{NV\textsuperscript{-}} upon laser irradiation \cite{leech2001}. The electron withdrawing character of halogenide, especially the fluorine, decreases the Fermi level of the nitrogen vacancy closer to its vacuum level, making loss of electron less plausible. Suprisingly, we have observed a switching of \textit{NV\textsuperscript{-}} to \textit{NV\textsuperscript{0}} upon surface termination. The effective surface termination was proven by X-ray photoelectron spectroscopy (XPS) and the presence of F1s and F KLL lines in the survey spectrum. Due to the insulating character of diamond and therefore the surface charging effects, the exact position of F1s and F KLL should not be taken as an absolute value, as it depends on the level of surface polarization \cite{CAZAUX1999155}:

\begin{equation}
    [\frac{-\hbar}{2m_{n}} \frac{d^{2}}{dz^{2}}+V_{z}]\psi_{i}(z)=\varepsilon_{i}\psi_{i}(z)  
\label{eq1}
\end{equation}

\begin{equation}
\frac{d^{2}\Phi }{dz}=-\frac{\rho (z)}{\epsilon}=-\frac{1}{\epsilon}2q[\sum_{mn}^{ } \sum_{i}f(\epsilon)\psi^{*}(z)] + e N_{0}=\frac{2q}{\epsilon}\sum_{i}N_{i}\psi^{2}_{i}(z+eN_{0})
\label{eq2}
\end{equation}
Here z is the term used to describe charge distance from the surface, $\psi_{i}$(z) is the i-th hole sub-band state wave function of eigenenergy $\epsilon_{i}$. $\Phi$(z) descibes the electrostatic potential, q is the charge, $\rho$(z) stands for the charge density, which is inversely proportional to the eigenenergy $\epsilon_{i}$.  Fluorination lowers the Fermi level of the nitrogen, making electron loss from successfully generated negatively charged nitrogen vacancy \textit{NV\textsuperscript{-}} thermodynamically favorable. According to the equation \ref{eq2}, as the Fermi level is lowered in energy, the charge density drops. We atribute this phenomenon to electron transfer from \textit{NV\textsuperscript{-}} to \textsuperscript{14}N\textsuperscript{+}. Many existing reports include possibilities of \textit{NV\textsuperscript{-}} activation by respective fluorine termination of the surface by means of plasma or wet chemical covalent (fluoride) and non-covalent (fluorine) termination \cite{nagl2015}, \cite{hu2013}, \cite{cui2013}. In contrast with these findings, we have observed selective deactivation of those color centers. This was perfomed by selective coverage of photonic structures by a quartz plate, of which most of the nanopillars were exposed to seemingly anisotropic SF\textsubscript{6} 0V plasma. The anisotropy of 0V plasma was surprisingly found to be anisotropic, as the free ion movement at 0V acceleration voltage in the plasma reactor would be expected to be governed only by Hamilton-Jacobi equations. Nevertheless, we were able to sufficiently cover the photonic structures with a quartz plate by a crude manner and still observe no underetching. We correlate this observation to the specifics of an ICP/RIE plasma reactor and for a detailed explanation, we refer the reader to our other manuscript dealing only with this particular topic. Here, we present an optical emission spectrum of this plasma showing the respective fluorine transitions being the source of effective surface termination. The respective transitions were assigned in Fig. \ref{fig:3}, which shows a possible mechanism of SF\textsubscript{6} decomposition. The charge state switching was found not to be reversible. Deactivation of the \textit{NV\textsuperscript{-}} centers was irreversible, but not judgdemental to the surface quality in terms of root mean square roughness (rms). This was proven by means of atomic force microscopy (AFM), whereby the increase in rooted mean square roughness (rms) did not increase nor decrease substantially from 1.5 nm regime. The actual influence was observed within photoluminescence countrate, which dropped by an order of magnitude after exposure to 0V bias SF\textsubscript{6} plasma. Although we have not observed any plasma-induced surface damage, in contrast with other researchers, we have employed SF\textsubscript{6} plasma, not the CHF\textsubscript{3} or CF\textsubscript{4}. The unique character of pure ICP-discharge in 0V SF\textsubscript{6} plasma, generated heavier ions even though they was found to etch the surface in a uniform fashion without causing the damage, substantially impaired though the stability of NV centers. We refer the fact of selective deactivation of negatively charged nitrogen vacancies to their thermodynamical instability \cite{drumm2010}. The nitrogen vacancies were not annealed and therefore even slight changes in the crystallographic surrounding induced by the bombardment with heavy ions like SF\textsubscript{5}\textsuperscript{+} cause a change to energetically more favorable neutral state. This assumption is well supported by the fact that large strain present in the diamond after CVD growth observed under optical confocal scanning microscopy (in range of 10E06 counts per second under 519 green laser irradiation in $\mu$W range). We therefore conclude that the naturally generated nitrogen vacancies are inferior to surface termination enhanced photon extraction. Similar procedures with fluoromethane plasma on diamond with \textsuperscript{14}N\textsuperscript{+} implanted by ion implantation techniques followed by high temperature high vacuum annealing has provided promising results, as well as the wet chemical fluorination \cite{nagl2015}. The fluorination induced charge state switching within naturally generated nitrogen vacancy, which was found to be irreversible. We attribute this effect to the presence of unreacted atomic nitrogen in the diamond as well as strain inducing physical etching with heavy SF$\textsubscript{6}$ plasma.
\section{Conclusions}
\label{sec:7}
In conclusion, we have developed a highly sophisticated photon emitter system based on naturally grown-in and stable nitrogen vacancies within single crystal diamond. We have successfully manipulated the spin of these naturally occuring negatively charged nitrogen vacancies with optically detected magnetometric techniques. These manipulations serve as a proof of concept for distinguishing other color centers present in the SCD. A broad distribution of this sample along the surface was correlated with measurements of \textit{NV\textsuperscript{-}} density. In addition we have observed that these color centers can be actively switched off by soft etching with 0V-biased fluorine plasma. This speaks well for the strong electron withdrawing effect of the halogenide attached to the surface (as proven by XPS) and instability upon stress relief induced by physical etching with heavy sulfur ions within the ICP RIE chamber. Interestingly, we have observed no signature of a negatively charged silicon vacancy, neither we witnessed any spectroscopic signature of other impurities present in diamond. 
\begin{acknowledgements}
We acknowledge Dr. Rene Hensel (INM, Germany) for granting access to the ICP RIE reactor. We note that the nanofabrication method applied in this study is filed for a patent, application number: EP19198772.6.
\end{acknowledgements}

 \section*{Conflict of interest}
 The authors declare that they have no conflict of interest.

 \bibliographystyle{siam}
  \bibliography{Refs}

\end{document}